\documentstyle[prl,aps,epsf]{revtex}
\begin{document}
\twocolumn[\hsize\textwidth\columnwidth\hsize \csname @twocolumnfalse\endcsname
\title{Magnetoresistance of Epitaxial Fe Wires with Varied
Domain Wall Structure}

\author{U. R\"{u}diger$^1$, J. Yu$^1$, S. S. P. Parkin$^2$ 
and A. D. Kent${^{1,*}}$}

\address{
$^1$ Department of Physics, New York University,\\
4 Washington Place, New York, New York 10003, USA}

\address{
$^2$ IBM Research Division, Almaden Research Center,
San Jose, California  95120, USA}

\date{June 6, 1998}

\maketitle

\begin{abstract}
The low temperature negative domain wall (DW) contribution to the resistivity
observed in expitaxial Fe microstructures has been investigated as a function
of film thickness. The DW spin structure changes from Bloch to Neel-like with
decreasing film thickness. Results suggest that an interplay between orbital
effects in the internal magnetic fields near DWs and thin film surface
scattering are at the origin of the observed negative wall contribution.
%\vspace{20 mm}

%\bf{Keywords:}
%Magnetoresistance, Magnetic Domains, Magnetic Force Microscopy
\end{abstract}
\label{firstpage}
]
\narrowtext

The effect of domain walls (DWs) on quantum and spin-dependent electron 
transport in magnetic nanostuctures has been the subject of a number of recent 
theoretical and experimental studies 
\cite{Kent,Ruediger,Gregg,Tatara,Otani,RuedigerAPL}. In particular, detailed
studies of the MR of expitaxial Fe wires with controlled stripe domains
revealed a negative DW contribution to the resistivity at low temperature
($<80$ K) \cite{Kent,Ruediger}. This remarkable result was difficult
to reconcile with existing theories of DW scattering \cite{Gregg}
and an interesting model based on weak localization phenomena \cite{Tatara}.
Here these effects are further studied in microfabricated wires of
varied thickness and, hence, DW spin structure.

\section{Fabrication and Magnetic Characterization}
Epitaxial Fe (110) films of 25 nm, 50 nm, 100 nm, and 200 nm thickness 
were grown on a-axis (11${\bar 2}$0) sapphire substrates \cite{Kent,Ruediger}.
From these films Fe wires of 2 $\mu$m linewidth were microfabricated with the
long axis perpendicular to the in-plane [001] uniaxial magnetocrystalline
easy axis.

Magnetic Force Microscopy (MFM) images in Fig.1 (a) - (c) show that the 
competition between magnetostatic, magnetocrystalline, and exchange
interactions leads to a stripe domain pattern. All images were performed
with a vertically magnetized tip in zero applied magnetic field after
in-plane magnetic saturation parallel to the wire axis. The MFM images
highlight the magnetic poles at the boundaries of the wire and above the
DWs. Fig. 1 (a) - (c) show that there is a transition of the DW type
as a function of thickness from Neel-like DWs (25 nm) to Bloch (50 to 100 nm)
to canted Bloch Walls (200 nm). The transition from a Neel 
to a Bloch wall occurs when the thickness of the film is in the range
of the DW width and is driven by the magnetostatic energy of the DW.
If the wire is thinner than the DW width, the shape anisotropy of the DW
favors a Neel-like DW in which the magnetization rotates in the plane.
Following the same arguments, if the film thickness is greater than
the DW width shape anisotropy favors Bloch-like DWs, with an out-of-plane
rotation of the magnetization between adjacent stripe 
domains. Additionally, the subdivision of the DWs into sections of opposite 
chirality lowers the magnetostatic energy, as seen in Fig. 1 (a)
and partly in (b). Canting of the walls as seen in Fig. 1 (c) further
reduces the magnetostatic energy.

\begin{figure}[htb]
%\epsfxsize=2.95in
%\centerline{\epsfbox{fig1.eps}}
\caption{ MFM images in zero applied field of a (a) 25 nm, 
(b) 100 nm, and (c) 200 nm thick Fe wire of 2 $\mu$m width 
after longitudinal saturation.}
\label{fig1}
\end{figure}

\section {Transport Properties}
Low Field MR measurements were performed with an in-plane applied field 
oriented either longitudinal ($\parallel$) or transverse ($\perp$) to 
the wire axis. The MR has the following general characteristics. 
There is structure to the MR in applied fields less than the saturation 
field ($H_s$), at which point the slope
changes, and the resistivity then increases monotonically with field. 
Interestingly, the resistivity at $H_s$ at high temperature
($T>70$ K) is larger 
in the longitudinal than in the transverse field orientation 
($\rho_{\perp}(H_s) > \rho_{\parallel}(H_s)$),
while at low temperatures
($T< 60$ K), the situation is reversed ($\rho_{\parallel}(H_s) >
\rho_{\perp}(H_s)$).

\begin{figure}[tbh]
\epsfxsize=8.5cm
\centerline{\epsfbox{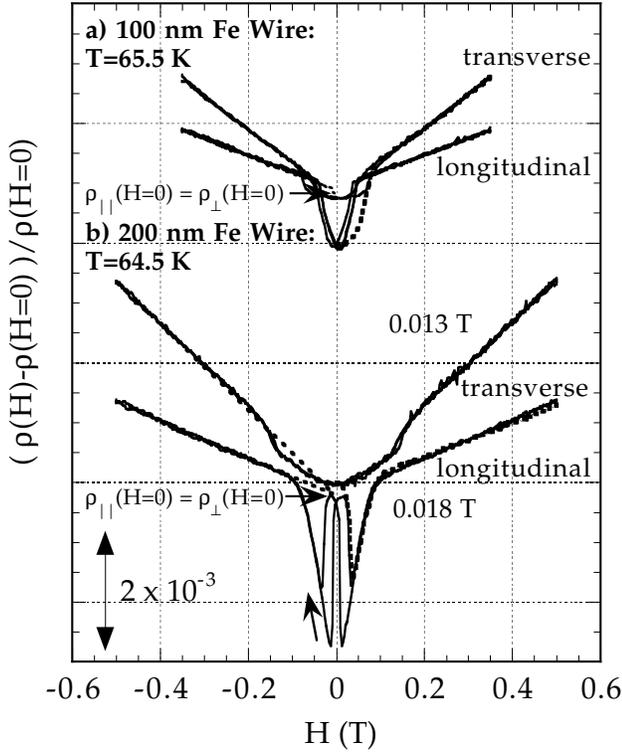}}
\caption{ MR data at $T=65.5 K$ of (a) 100 nm and (b) 
200 nm thick 2 $\mu$m linewidth Fe wires. The extrapolation 
of the high field data shows  $\rho_\perp (H=0)=\rho_\parallel(H=0)$, 
i.e. the absence of resistivity anisotropy.}
\label{fig2}
\end{figure}

In these wires there are two predominant and competing sources of resistivity 
anisotropy. The first has its origins in the spin-orbit interaction and 
is know as anisotropic MR (AMR)--the resistivity extrapolated back to zero 
internal field ($B=0$) depends on the angle between ${\bf M}$  and ${\bf J}$. 
The second effect is due to the anisotropy of the Lorentz MR, which 
generally depends on the angle of {\bf J}  and {\bf B}. 
Detailed MR measurements as a function of temperature
and field angle have been used to determine these anisotropy contributions 
\cite{Ruediger}. At low temperatures, because of the 
large internal field in Fe ($4\pi M=2.2$ T) and the decreased electron 
scattering rate, the anisotropy in the Lorentz MR is dominant, and 
leads to the observed reversal of the resistivity anisotropy.

For the purposes of this study it is sufficient to note that 
due to these competing effects there will be a temperature at 
which the in-plane resistivity anisotropy vanishes. We call this 
the compensation temperature $T_{comp}$ and it is defined as the 
temperature at which
$\rho_{\parallel}(H=0,T_{comp})=\rho_{\perp}(H=0,T_{comp})$.
This occurs close to $65 K$ for 
all the wires we have studied. Fig. 2 shows MR results 
at $T_{comp}$ for wires of 
different thickness ((a) 100 nm and (b) 200 nm). The slope in the MR above the 
saturation field is due to the Lorentz effect and the extrapolation of the 
high field MR to $H=0$ (dashed lines) illustrates the resistivity 
compensation. 

At this temperature the low field MR due to the magnetization 
reversal of the wires and, particularly, the in-plane reorientation 
of the magnetization should vanish. Fig. 2 (a) shows that a 
positive MR is associated with the erasure of DWs in the 100 nm 
thick wire, and is largest in the longitudinal field geometry in 
which the DW density at H=0 is greatest \cite{Ruediger}. 
These results have been taken as evidence for a negative DW 
contribution to the resistivity. The DW MR is calculated as,
$\rho_d=\rho(H=0)-\rho_{eff}(H=0)$, 
where $\rho(H=0)$ is the resistivity measured at $H=0$ and $\rho_{eff}(H=0)$,
is the extrapolation to H=0. The magnitude of $\rho_d$ increases with 
increasing DW density \cite{Ruediger,RuedigerAPL}

Fig. 2 (b) shows the MR of a 200 nm thick wire. The
longitudinal MR is now negative at low fields and positive at 
higher fields. This low field negative MR is 
associated with a change in the magnetization reversal mode
in thicker wires.  The insets to Fig. 2 show MFM 
images of the magnetic structure in longitudinally applied fields at room 
temperature. In the 100 nm wire the reversal proceeds via the growth of 
favorably oriented longitudinally magnetized in-plane closure domains (inset 
Fig. 2(a)), 
whereas for the 200 nm thick film stripe domain configurations are observed
(inset Fig. 2(b)). This latter image suggests
that the magnetization has rotated out 
of the plane along \{100\} easy directions 45 degrees to the surface normal. 
The reduction in internal field (due to the strong demagnetization fields 
perpendicular to the film plane) is  responsible for the low field 
drop in resistivity. 

By establishing the magnetic state shown in Fig. 1(c) at room temperature
by appropriate sample demagnetization and then
cooling the sample to the MR measurement temperature (64.5 K), we find 
that the local maximum in the resistivity (longitudinal curve Fig. 2(b))
at $H=0$
is associated with the canted DW structure (seen in Fig. 1(c)). 
Since the measured $H=0$ resistivity is now observed to equal to the $H=0$
resistivity
extrapolated from the high field MR, $\rho_d$ vanishes at this temperature 
for this wire.

\begin{figure}[htb]
{\small TABLE 1. Characteristic data for 2 $\mu$m linewidth wires as a
function of thickness.\\ \\}
\epsfxsize=2.95in
\centerline{\epsfbox{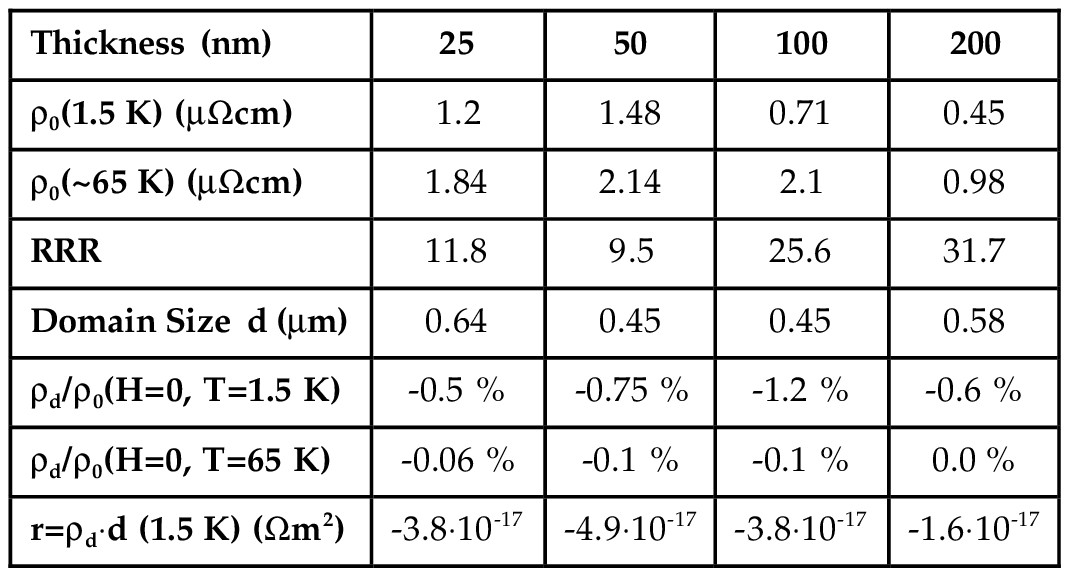}}
\end{figure}

Varying the film thickness changes both the DW spin structure (Fig. 1) and the 
relative importance of bulk and thin film surface scattering on the 
resistivity. Table 1 summarizes the results of a systematic 
study of the effect of film thickness on $\rho_d$ and the film resistivity.  
It is seen that even in thin layers (25 nm films), in which the 
DWs are considerably broader and Neel-like, there is a large anomalous 
negative DW contribution. In fact, the magnitude of the DW 
interface resistance (which is negative) and given by $r=\rho_dd$  
where $d$ is the domain size, is largest in the thinner films ($\leq 100$ nm). 

\begin{figure}[htb]
\epsfxsize=2.95in
\centerline{\epsfbox{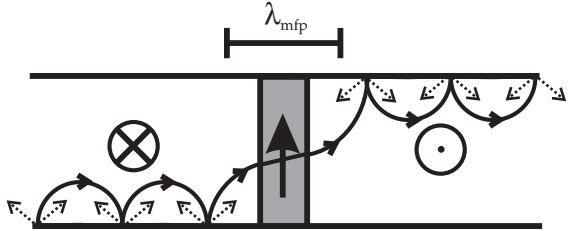}}
\caption{Cross-sectional view of the magnetic configuration of an Fe wire,
showing the effect of internal fields and surface 
scattering on the trajectory of charge carriers
within stripe domains and DWs.}
\label{fig3}
\end{figure}

These results suggest that the interplay between orbital 
effects in the internal 
magnetic fields near DWs and surface scattering
are at the origin of the anomalous negative wall contribution. 
Fig. 3 illustrates a manner in which the trajectories of 
charge in the alternating internal magnetic field near a 
wall may lead to reduced surface interaction and hence an 
enhanced conductivity. Demagnetization fields in the wall 
will also act to increase the conductivity via the Lorentz effect. 
However, in thicker films, in which bulk scattering is dominant, 
we find the $r$ decreases. Altering the wall size and 
structure--which changes the demagnetization fields near 
walls--also is not critical to the observed effects. 
It appears that the magnetic structure within approximately 
a mean free path, $\lambda_{mfp}$ ($\sim 200$ nm at 1.5 K in a 100 nm 
thick film) is important to the observed phenomena.

This research was supported by DARPA-ONR, Grant \# N00014-96-1-1207. 
Microstructures were prepared at the CNF, project \# 588-96.
\\ \\ {\small  $^*$Electronic address: andy.kent@nyu.edu}

\end{document}